\documentclass[journal=jctcce,manuscript=article]{achemso}
\SectionNumbersOn

\usepackage{graphicx} 
\usepackage{amsmath, bm, amsfonts} 
\usepackage{braket}
\usepackage{xcolor}
\usepackage{tabularx}
\usepackage{multicol,multirow}
\usepackage[ruled, noline]{algorithm2e} 
\definecolor{mitred}{HTML}{750014}
\definecolor{mitgray}{HTML}{8b959e}

\usepackage{duckuments}

\usepackage{xr}

\title{Selected Coupled Cluster Guided by Selected Configuration Interaction}

\author{Minsik Cho}
\affiliation[MIT]{Department of Chemistry, Massachusetts Institute of Technology, Cambridge, MA 02139, USA}

\author{Troy Van Voorhis}
\email{tvan@mit.edu}
\affiliation[MIT]{Department of Chemistry, Massachusetts Institute of Technology, Cambridge, MA 02139, USA}

\begin{document}

\maketitle

\begin{abstract}
Coupled Cluster (CC) theory is one of the most popular correlated wavefunction methods, yet the steep computational cost scaling of CC truncated at higher rank limits its application. Modern Selected Configuration Interaction (SCI) methods successfully exploit the sparsity of ground-state wavefunctions in the determinant space, motivating an analogous approach with the exponential CC ansatz. In this work, we introduce Selected Coupled Cluster (SCC), an arbitrary-order sparse CC method that explicitly solves the CC residual equations within a compact amplitude space guided by an SCI wavefunction. SCC optimizes the CC amplitudes in the sparse amplitude space. Using a pilot Just-in-Time (JIT) compiled implementation, we benchmark SCC against SCI, SCI+PT2, standard rank-truncated dense CC (CCSD, CCSDT), and DMRG across various chemical systems. We demonstrate that SCC exhibits fast, monotonic convergence with the SCI variational threshold, outperforming SCI and surpassing SCI+PT2 in the tight-threshold regime. SCC also provides accurate energies for high-dimensional topologies, where DMRG typically converges slowly with bond dimension. SCC thus presents an efficient drop-in enhancement for modern SCI workflows.
\end{abstract}

\section{Introduction}\label{sec:intro}

Coupled cluster (CC) theory has established itself as one of the most popular choices in high-accuracy simulation of various electronic structure problems. \cite{cc1,cc2,cc3,cc4} In particular, CCSD(T) -- coupled cluster singles, doubles, and perturbative triples -- is often referred to as the \textit{gold standard} that routinely captures a significant portion of electron correlation effects and accurately models various molecular properties.\cite{ccsd_t1,ccsd_t2} Higher-order CC calculations, such as CCSDT and CCSDTQ, include higher-rank excitation operators and yield systematically improvable accuracy.\cite{ccsdt,ccsdtq} These extremely high-quality CC calculations are possible only for a very limited range of system sizes due to their steep computational cost scaling. CCSDT and CCSDTQ, for example, has $\mathcal{O}(N_o^3 N_v^5)$ and $\mathcal{O}(N_o^4 N_v^6)$ scalings respectively, where $N_o$ and $N_v$ are the number of occupied and virtual orbitals. This steep scaling originates from the number of wavefunction parameters required to represent the higher-order cluster operators. Physical intuition and empirical observations, such as the success of various local CC schemes,\cite{dlpno_ccsd,lpno_ccsd} suggest that the vast majority of these amplitudes are negligibly small. This sparsity of CC wavefunctions implies an opportunity for significant computational cost savings compared to conventional implementations that utilize dense linear algebra routines.

Exploiting the sparsity of ground-state wavefunctions, as in a sparse set of amplitudes, has been a popular direction explored in a few earlier works. Adaptive Coupled Cluster (@CC) constructs orbital importance functions from a CISD wavefunction and optimizes a CC wavefunction within a selected set of arbitrary-rank amplitudes.\cite{atcc} Full Coupled-Cluster Reduction (FCCR) similarly treats sparse arbitrary-rank amplitudes, but instead expands the amplitude space rank-by-rank.\cite{fccr} Single-commutator terms are evaluated to identify the significant higher-rank amplitudes connected to the previously chosen amplitudes. In Unitary Selected Coupled Cluster (USCC), the product of Hamiltonian matrix elements and $T$-amplitudes identifies the important singles and doubles for a variational quantum eigensolver (VQE) ansatz.\cite{uscc} Triple and quadruple amplitudes are then iteratively selected by thresholding similar products of Hamiltonian matrix elements and $T$-amplitudes. Coupled Cluster Monte Carlo (CCMC) performs a stochastic algorithm that evolves the population of discretized fictitious particles called excips in imaginary time to estimate the important amplitudes.\cite{ccmc} Finally, we note various approaches that select at the level of a subspace or an active space rather than individual amplitudes. That is, they include higher-order amplitudes only within a particular subspace rather than including only some amplitudes in all spaces. Notable examples include Tailored Coupled Cluster (TCC)\cite{tcc-casci,tcc-dmrg,tcc-benchmark} and CC$(P; Q)$.\cite{ccpq1,ccpq2,ccpq3,ccpq4} In these approaches, only some amplitudes are fully optimized with CC residual expressions. Others are either frozen or approximated.

Analogous approaches that utilize the sparsity of ground-state wavefunctions exist in the configuration interaction (CI) literature, with notable successes in recent contributions. These Selected CI (SCI) methods generally work with a truncated list of determinants, chosen based on affordable estimates of their significance to the ground-state wavefunction. An early example of SCI is Configuration Interaction using a Perturbative Selection made Iteratively (CIPSI).\cite{cipsi} CIPSI iteratively expands the determinant space by evaluating the second-order Epstein-Nesbet perturbation for all external determinants. Calculating the perturbative contributions in CIPSI requires substantial resources, and modern SCI methods have gained traction thanks to more efficient schemes to identify the determinant subspace. Notable examples are Heat-bath CI (HCI)\cite{arrow3} and Semistochastic Heat-bath CI (SHCI)\cite{arrow1,arrow2}, which use the product between Hamiltonian matrix elements and CI coefficients to expand the determinant space. The computationally efficient selection rule, combined with robust perturbative treatment of the external space, attracted interest in this modern SCI method. Related noteworthy methods in this direction include Full Configuration Interaction Quantum Monte Carlo (FCIQMC) \cite{fciqmc}, Adaptive CI (ACI) \cite{aci}, and Adaptive Sampling CI (ASCI) \cite{asci}.

In this paper, we show that the efficient selection schemes created for SCI can be used to efficiently define amplitude space for highly accurate, arbitrary order SCC calculations. Unlike typical subspace-based approaches, SCC explicitly optimizes all selected CC amplitudes simultaneously; the SCI wavefunction only informs the amplitude space of the CC wavefunction. SCC resembles FCCR and @CC in that the amplitude choices are guided, yet differs in that a full-space modern SCI wavefunction drives the amplitude selection. The method aims to benefit from the robust configuration selection scheme that powers modern SCI methods. We show that SCC performed in this manner exhibits fast and monotonic convergence to the exact ground state energies with SCI threshold $\varepsilon_\text{var}$. We benchmark SCC against SCI, SCI+PT2, and DMRG, for a range of chemical systems, including a polycyclic aromatic hydrocarbon (PAHs), a metal complex, and an inorganic cluster compound. We observe that SCC energies are much more accurate than SCI energies for the same memory resources. SCC also outperforms SCI with perturbative corrections (SCI+PT2) in the tighter variational threshold regime. SCC, in particular, provides robust energy estimates for systems with a high-dimensional topology, where prohibitively large bond dimensions are required to obtain accurate DMRG energies. For a comparable number of amplitudes, SCC also surpasses the accuracy of standard rank-truncated CC calculations, such as CCSD and CCSDT.

In Section \ref{sec:theory}, we first review the working equations of CC and cluster decomposition of CI wavefunctions. We then explain how we utilize Just-in-Time (JIT) compilation to implement arbitrary-order sparse CC. We also point to how the implementation employs existing reusable libraries, such as \texttt{wick\&d}\cite{wicked} and \texttt{clusterdec}\cite{clusterdec}. In Section \ref{sec:comp_details}, we specify the computational details of the benchmark calculations, which follow in Section \ref{sec:results}. We finally discuss the implications of the results in Section \ref{sec:discussion} and conclude in Section \ref{sec:conclusion} with future directions that naturally arise from this work.

\section{Theory}\label{sec:theory}

\subsection{Coupled Cluster Theory}
We first briefly introduce CC theory. For a more in-depth discussion of the CC method in quantum chemistry, we refer readers to relevant reviews.\cite{cc4,cc_review}

The fundamental distinction of CC theory, compared to configuration interaction (CI), lies in the choice of an exponential wavefunction ansatz. For a chemical system, we start with a time-independent Schr\"{o}dinger Equation
\begin{equation}\label{eq:schrodinger}
    \hat{H} \ket{\psi} = E \ket{\psi}
\end{equation}.
$\hat{H}$ is the Hamiltonian operator of the system, and $\ket{\psi}$ is an exact wavefunction associated with an energy $E$. In CC theory, we express the wavefunction $\ket{\psi}$ with the following ansatz:
\begin{equation}
    \ket{\psi} = e^{\hat{T}}\ket{\phi_0}
\end{equation}
$\hat{T}$ is the cluster operator, and $\ket{\phi_0}$ is some reference wavefunction, typically the Hartree-Fock determinant. The cluster operator is a sum of excitation operators that move electrons from occupied reference orbitals to unoccupied orbitals. These are typically denoted with the excitation rank $k$, which specifies the number of electrons relocated from the reference determinant $\ket{\phi_0}$.
\begin{equation}
    \hat{T} = \sum_k \hat{T}_k = \frac{1}{(k!)^2} \sum_{o_1,o_2,\cdots,o_k} \sum_{v_1,v_2,\cdots,v_k} t_{o_1 o_2 \cdots o_k}^{v_1 v_2 \cdots v_k} \hat{a}_{v_1}^\dagger \cdots \hat{a}_{v_k}^\dagger \hat{a}_{o_k} \cdots \hat{a}_{o_1}
    \label{eq:tamp}
\end{equation}
Here, $o$ and $v$ label the occupied and unoccupied orbitals, respectively. $t_{o_1 o_2 \cdots o_k}^{v_1 v_2 \cdots v_k}$ is the amplitude associated with the particular excitation from the reference wavefunction $\ket{\phi_0}$. CC calculations obtain the ground state solution by optimizing these wavefunction parameters. The CC working equations are based on projection onto the reference and excited determinants and are as follows:
\begin{align}
    \bra{\phi_0}e^{-\hat{T}}\hat{H}e^{\hat{T}} \ket{\phi_0} &= E \label{eq:ccenergy} \\
    R_\mu := \bra{\mu}e^{-\hat{T}}\hat{H}e^{\hat{T}} \ket{\phi_0} &= 0 \label{eq:residual}
\end{align}
where $\ket{\mu}$ is some excited determinant such as $\ket{\phi_{i}^{a}}, \ket{\phi_{ij}^{ab}}, \cdots$. Equation \ref{eq:ccenergy} allows evaluation of the energy once the CC amplitudes are determined, and Equation \ref{eq:residual}, or the residual expression, is the loss function optimized to locate the optimal CC amplitudes. The similarity-transformed Hamiltonian $e^{-\hat{T}}\hat{H}e^{\hat{T}}$ can be expressed as nested commutators by applying the Baker-Campbell-Hausdorff expansion and can be expanded using computer algebra systems by applying Wick’s theorem.\cite{wicktheorem}

Standard CC implementations truncate the list of excitation operators by the rank $k$. Including higher-rank excitation operators increases computational complexity and storage cost, while improving the energy estimates. This trade-off naturally offers a systematically improvable workflow, from CC Singles and Doubles (CCSD) to higher-rank approximations like CC Singles, Doubles, and Triples (CCSDT) and CC Singles, Doubles, Triples, and Quadruples (CCSDTQ).

While the rank-truncated CC provides robust control over computational resources and accuracy, not all individual amplitudes contribute significantly to the ground-state wavefunction. Analogous to how SCI exploits this sparsity by limiting to a chosen list of configurations, SCC operates within a list of sparse, arbitrary-rank CC amplitude space. In other words, SCC treats a compact list of the T amplitudes in Equation \ref{eq:tamp} regardless of the rank. Unlike existing works in the literature, such as FCCR and @CC, that propose various selection criteria, we do not make a bespoke amplitude-selection criterion. Instead, we employ a modern SCI, Heat-bath CI, to guide our choice of the CC ansatz and focus on the characteristics of sparse arbitrary-rank CC calculations in this work.

\subsection{Cluster Decomposition of CI wavefunctions}\label{sec:clusterdec}

To use the SCI wavefunction as a guide for SCC, it is necessary to convert the CI coefficients into equivalent CC amplitudes. We use the cluster decomposition procedure implemented by Lehtola and coworkers\cite{clusterdec}, applicable to general-order CC amplitudes. For completeness, we present here an abridged explanation of the workflow.

The full, non-truncated, CI and CC wavefunctions are equivalent -- they only differ in how the wavefunction is parametrized. This suggests an equality between the two representations
\begin{align}
    \ket{\psi} &= \left(1 + \hat{C}\right) \ket{\phi_0} =  \left(1 + \hat{C}_1 + \hat{C}_2 + \hat{C}_3 + \hat{C}_4 + \cdots\right)\ket{\phi_0} \\
    &= e^{\hat{T}} \ket{\phi_0} =  \exp{\left(\hat{T}_1 + \hat{T}_2 + \hat{T}_3 + \hat{T}_4 + \cdots\right)}\ket{\phi_0}
\end{align}
where $\hat{C}_n$ and $\hat{T}_n$ represent the CI coefficients of the rank-$n$ excited determinants and the rank-$n$ cluster operators, respectively. $\ket{\phi_0}$ denotes the reference wavefunction, often chosen as the Hartree-Fock wavefunction. We can expand the exponential operator using a Taylor series.
\begin{equation}
    \exp{\left(\hat{T}_1 + \hat{T}_2 + \hat{T}_3 + \hat{T}_4 + \cdots\right)} = 1 + \hat{T} + \frac{1}{2!} \hat{T}^2 + \frac{1}{3!} \hat{T}^3 + \cdots
\end{equation}
Collecting the terms rank-by-rank, the relationship between the CI and CC operators for the first three ranks follow:
\begin{align}
    \hat{T}_1 &= \hat{C}_1 \\
    \hat{T}_2 &= \hat{C}_2 - \frac{1}{2}\hat{T}_1^2 \label{eq:c2}\\
    \hat{T}_3 &= \hat{C}_3 - \frac{1}{2}\left(\hat{T}_1\hat{T}_2 + \hat{T}_2\hat{T}_1\right) - \frac{1}{3!} \hat{T}_1^3
\end{align}
We calculate the equivalent CC amplitudes for given CI coefficients using a cluster decomposition implementation \texttt{clusterdec}.\cite{wicktheorem,clusterdec} From a list of sparse CI coefficients, \texttt{clusterdec} automatically generates the Wick contractions and evaluates the equivalent CC representation. Starting from a list of sparse CI coefficients, we can then identify the significant CC amplitudes.

\subsection{Arbitrary-order Sparse Coupled Cluster Implementation}
Unlike rank-truncated CC implementations that operate on a dense set of amplitudes, an arbitrary-order sparse coupled-cluster code requires specialized software design to keep storage requirements low and scale only with the number of selected amplitudes. To this end, as we describe in more detail, we employed a sparse data structure and a Just-in-Time (JIT) transpiler that outputs C++ code to materialize a pilot implementation efficient enough to assess the SCC approach. The code, which we maintain on a public repository\cite{qscc_repo}, uses \texttt{PySCF}\cite{pyscf} to evaluate the molecular integrals and calls \texttt{wick\&d}\cite{wicked} to derive the symbolic residual expressions.
To illustrate the workflow, we consider an example of a term that contributes to the $R_2$ residual that we obtain by calling \texttt{wick\&d} in the code:
\begin{equation}
    R_{ij}^{ab} \leftarrow \sum_m F_{im} t_{mj}^{ab}
\end{equation}
Here, $F$ denotes the one-body Fock matrix. To start, the residual tensor $R_{ij}^{ab}$ is sparse. The tensor contraction routine, therefore, iterates only over the selected \textit{active} indices. This outer loop is parallelized using OpenMP to evaluate the independent residual contributions concurrently. For a fixed set of active indices $(i, j, a, b)$, the remaining task is to sum over the index $m$, noting that $F_{im}$ is dense and $t_{mj}^{ab}$ is sparse. For optimal performance, we use a Compressed Sparse Row (CSR) data structure with a cache-friendly hash table for the $T$-amplitudes. Specifically, for a $T$-amplitude tensor of rank $k$, the sparse matrix represents the hole index tuples as rows and the particle index tuples as columns. The hash table takes a hole index as an input and returns the corresponding active particle indices and the contiguous memory location.

To further minimize the lookup costs across high-dimensional tensors, the JIT transpiler groups residual terms that share common data lookups and writes separate C++ functions for each. The generated C++ code is compiled at runtime, enabling various compiler optimizations, such as loop unrolling and vectorization, tailored to the hardware. The numerical studies for this work, for example, were conducted on an in-house computing cluster equipped with Intel Xeon Gold 5220R CPUs and leveraged fused multiply-add operations and the AVX-512 vector instruction set.

\subsection{Selected Coupled Cluster Workflow}

\begin{figure}[tb]
    \centering
    \includegraphics[width=0.5\linewidth]{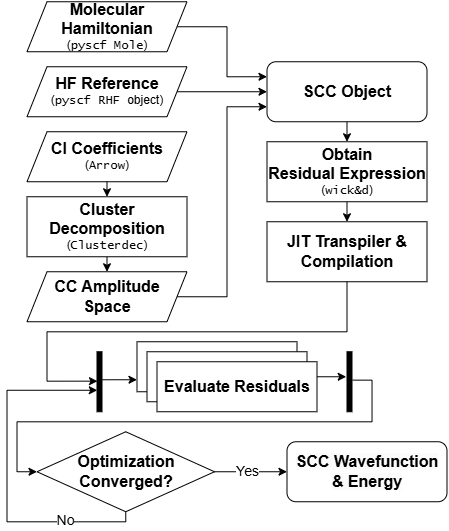}
    \caption{Flowchart of SCC Implementation}
    \label{fig:flowchart}
\end{figure}

Figure \ref{fig:flowchart} illustrates the combined workflow for SCC calculations. \texttt{PySCF}\cite{pyscf} provides the Hartree-Fock reference and the molecular Hamiltonian integrals. We convert the sparse set of CI coefficients from \texttt{Arrow}\cite{arrow1,arrow2,arrow3} to CC amplitudes with \texttt{clusterdec}\cite{clusterdec} to define the CC amplitude subspace. Our SCC implementation calls \texttt{wick\&d}\cite{wicked} to obtain the CC residual expressions up to the maximum rank in the chosen CC amplitude subspace. Our code then writes C++ code that evaluates the residuals term by term with sparse storage. Convergence algorithms, such as DIIS, reside in the Python part, and the compute-intensive residual evaluation is delegated to the efficient compiler-optimized C++ code.

\section{Computational Details}\label{sec:comp_details}

Benchmark calculations were performed on a selection of chemical systems to assess SCC. Specifically, we ran SCI (Heat-bath CI without perturbative correction), SCI-PT2 (Semistochastic Heat-bath CI), SCC, CCSD, CCSDT, and DMRG on the target systems. To cover a diverse range of chemical systems, we considered i) butane, ii) circumcoronene, iii) zinc phthalocyanine (ZnPc), and iv) boron cluster (dodecaborate anion; $[\text{B}_{12}\text{H}_{12}]^{2-}$). The number of electrons and orbitals for each system, as well as the active space construction (if any), are specified in Table \ref{table:as}. For calculations on active spaces, we utilized AVAS to construct the appropriate space from the canonical HF molecular orbitals.\cite{avas1,avas2} We performed the SCC calculations using the publicly available pilot implementation described in the previous section.\cite{qscc_repo} Each SCC calculation was converged until the norm of the residual vector reached less than $10^{-6}$. SCI and SCI-PT2 calculations were conducted with Arrow to obtain the Heat-bath CI results.\cite{arrow1,arrow2,arrow3} Variational threshold ($\varepsilon_\text{var}$) was varied as illustrated in Section \ref{sec:results}, and the perturbative calculation included determinants according to the default setting. ($\varepsilon_\text{PT, deterministic} = \max(\varepsilon_\text{var}^2 / 10, 2\times 10^{-6})$, $\varepsilon_\text{PT, stochastic} = \max(\varepsilon_\text{var}^2 / 100, 1\times 10^{-7})$, and $\varepsilon_\text{PT} = \varepsilon_\text{var}^2 / 1000$) \texttt{Clusterdec} was used to identify the CC amplitude subspace from the Heat-bath CI wavefunction, following the cluster decomposition steps in Section \ref{sec:clusterdec}. We report in Section \ref{sec:results} how many CC amplitudes we select from the full CC amplitude space, along with the numerical results, for easy comparison. The current implementation takes a restricted HF reference and obtains spin-orbital-based residual equations from \texttt{wick\&d}. The total number $N_k$ of unique, anti-symmetrized, and spin-conserving cluster operators in the spin-orbital basis at rank $k$ is:
\begin{equation}
  N_k = \sum_{m=0}^k \binom{n_{\text{occ}}}{m} \binom{n_{\text{occ}}}{k-m} \binom{n_{\text{vir}}}{m} \binom{n_{\text{vir}}}{k-m}\end{equation}
where $n_\text{occ}$ and $n_\text{vir}$ are occupied and virtual spatial orbitals respectively. $m$ denotes the number of $\alpha$ excitations, such that $k-m$ represents the number of substituted $\beta$ electrons. Future implementation could incorporate spin-free formalism for restricted SCC calculation to improve computational efficiency. We calculated the CCSD and CCSDT energies using \texttt{PySCF}\cite{pyscf,pyscf_ccsdtq}. DMRG results were obtained using \texttt{block2} with Boys-localized orbitals and Fiedler orbital ordering.\cite{block2-1,block2-2,block2-3} Extrapolation of DMRG results to the infinite bond dimension followed a standard approach based on linear extrapolation of the discarded weights.

\begin{table}[tb]
    \centering
    \begin{tabular}{c|c|c|c}
        System & Orbital Space & Active Space Selection & Basis Set \\ 
        \hline
        Butane         & (34e, 30o) & Full space & STO-3G \\
        Circumcoronene & (54e, 54o) & C $2p_z$ & def2-SVP \\
        ZnPc           & (52e, 45o) & C $2p_z$, N $2p_z$, Zn $3d$ & def2-SVP \\
        Boron cluster  & (72e, 72o) & Full space & STO-3G \\
    \end{tabular}
    \caption{System size and active space construction for the benchmark systems}
    \label{table:as}
\end{table}

\section{Results}\label{sec:results}

\subsection{Butane}
\begin{figure}[tb]
    \centering

    \centering
    \includegraphics[width=0.7\linewidth]{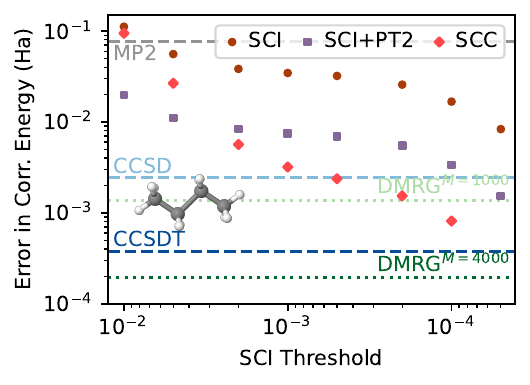}

    \resizebox{\linewidth}{!}{
    \setlength{\tabcolsep}{4pt}
    \begin{tabular}{c|c|*{6}{>{\centering\arraybackslash}p{1.6cm}}| >{\centering\arraybackslash}p{1.6cm}}
        \hline
        SCI & SCC Corr.& \multicolumn{7}{c}{Number of Amplitudes}\\ \cline{3-9}
        Threshold & Energy (Ha) & $T_1$ & $T_2$ & $T_3$ & $T_4$ & $T_5$ & $T_6$ & Total \\
        \hline
        $1\times10^{-2}$ & -0.203886 & 0 & 1,471 & 0 & 0 & 0 & 0 & 1,471\\
        $5\times10^{-3}$ & -0.271628 & 0 & 4,153 & 0 & 0 & 0 & 0 & 4,153\\
        $2\times10^{-3}$ & -0.292637 & 2 & 8,609 & 48 & 147 & 0 & 0 & 8,806\\
        $1\times10^{-3}$ & -0.295093 & 12 & 11,527 & 432 & 1,491 & 0 & 0 & 13,462\\
        $5\times10^{-4}$ & -0.295918 & 74 & 13,863 & 2,856 & 5,796 & 0 & 0 & 22,589\\
        $2\times10^{-4}$ & -0.296749 & 96 & 16,111 & 39,462 & 77,302 & 6 & 0 & 132,977\\
        $1\times10^{-4}$ & -0.297472 & 102 & 17,089 & 179,336 & 510,664 & 406 & 91 & 707,688\\ \hline
        0 (Full) & & 442 & 70,057 & $5\times10^{6}$ & $2\times10^{8}$ & $5\times10^{9}$ & $8\times10^{10}$\\
        \hline
    \end{tabular}
    }
    
    \caption{(Top) Error in correlation energy relative to extrapolated DMRG for butane\\
    (Bottom) SCC amplitude count (in spin orbitals) and raw correlation energies
    }
    \label{fig:butane}
\end{figure}

We first discuss the benchmark results on a minimal-basis butane. This is a small, easy system where we expect the electron correlation to be weak. Figure \ref{fig:butane} shows the error in correlation energies for the benchmarked methods, relative to the extrapolated DMRG energy. For this quasi-linear molecule, we observe rapid convergence of DMRG results with bond dimension and therefore use the extrapolated DMRG energy as our best estimate of the reference energy. We also observe that SCC always provides more accurate energies than the SCI calculation at the same threshold. SCC is less accurate than SCI-PT2 in the very-loose SCI-threshold regime, but converges to the ground-state energy much faster, surpassing SCI-PT2’s accuracy in the tighter regime. We also note that the SCC calculation of a comparable number of amplitudes surpasses CCSD in accuracy. We report the raw correlation energies in the Supporting Information for reference.

\subsection{Circumcoronene}

We then benchmark the circumcoronene system, where electron correlation is expected to be less trivial compared to butane. Due to the two-dimensional topology, DMRG is expected to converge more slowly as the bond dimension increases. SCC and SCI(+PT2) would also require a larger amplitude (configuration) space that captures the stronger correlation. Thus, of the traditional approaches to our disposal, we expect CCSDT to provide the most accurate correlation energy for this system and therefore use the CCSDT energy to guide extrapolative schemes. In the Supporting Information, we extrapolate the SCC and SCI+PT2 correlation energies. Generally, we find that the extrapolated SCI+PT2 energy with the accessible SCI threshold $\varepsilon_\text{var}$ is not better than the CCSDT energy.

\begin{figure}[tb]
    \centering
    \includegraphics[width=0.7\linewidth]{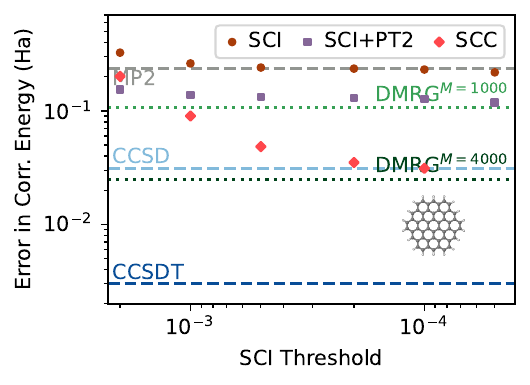}

    \resizebox{\linewidth}{!}{
    \setlength{\tabcolsep}{4pt}
    \begin{tabular}{c|c|*{6}{>{\centering\arraybackslash}p{1.6cm}}| >{\centering\arraybackslash}p{1.6cm}}
        \hline
        SCI & SCC Corr.& \multicolumn{7}{c}{Number of Amplitudes}\\ \cline{3-9}
        Threshold & Energy (Ha) & $T_1$ & $T_2$ & $T_3$ & $T_4$ & $T_5$ & $T_6$ & Total \\
        \hline
        $2\times10^{-3}$ & -0.438530 & 0 & 31,607 & 0 & 0 & 0 & 0 & 31,607\\
        $1\times10^{-3}$ & -0.549834 & 0 & 70,717 & 0 & 0 & 0 & 0 & 70,717\\
        $5\times10^{-4}$ & -0.591542 & 4 & 111,211 & 20 & 9 & 0 & 0 & 111,244\\
        $2\times10^{-4}$ & -0.604909 & 54 & 146,125 & 2,124 & 2,417 & 0 & 0 & 150,720\\
        $1\times10^{-4}$ & -0.608761 & 168 & 160,729 & 48,156 & 112,354 & 0 & 0 & 321,407\\
        $5\times10^{-5}$ &  & 326 & 173,305 & 560,730 & 1,800,220 & 124 & 157 & 2,534,862\\ \hline
        0 (Full) & & 1,458 & 777,843 & $2\times10^{8}$ & $3\times10^{10}$ & $3\times10^{12}$ & $2\times10^{14}$\\
        \hline
    \end{tabular}
    }
    
    \caption{(Top) Error in correlation energy relative to extrapolated SCC energy for circumcoronene
    \\
    (Bottom) SCC amplitude count (in spin orbitals) and raw correlation energies
    }
    \label{fig:circum}
\end{figure}

We find that SCC exhibits several properties observed in the butane system. SCC shows robust convergence with the SCI threshold and provides more accurate correlation energies than SCI with identical SCI thresholds. Except for the calculation using the loosest SCI threshold, SCC provides more accurate energies than SCI+PT2. As noted in Figure \ref{fig:circum}, the SCC calculation with the tightest SCI threshold $\varepsilon_\text{var} = 10^{-4}$ optimizes fewer amplitudes than the CCSD calculation yet evaluates comparable energies. We present the full raw data in the Supporting Information.

\subsection{Stretched Circumcoronene}

\begin{figure}[tb]
    \centering
    \includegraphics[width=0.7\linewidth]{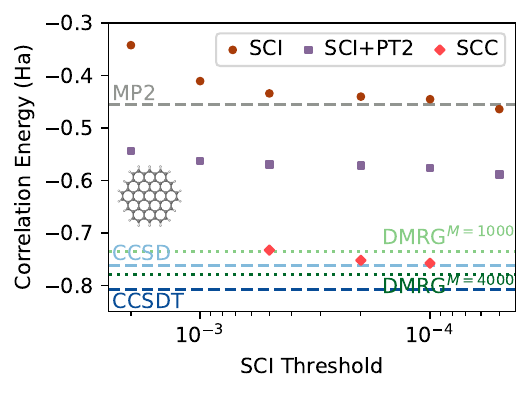}

    \setlength{\tabcolsep}{4pt}
    \begin{tabular}{c|c|*{4}{>{\centering\arraybackslash}p{1.6cm}}| >{\centering\arraybackslash}p{1.6cm}}
        \hline
        SCI & SCC Corr.& \multicolumn{5}{c}{Number of Amplitudes}\\ \cline{3-7}
        Threshold & Energy (Ha) & $T_1$ & $T_2$ & $T_3$ & $T_4$ & Total \\
        \hline
        $2\times10^{-3}$ &  & 0 & 31,691 & 0 & 0 & 31,691\\
        $1\times10^{-3}$ &  & 0 & 68,993 & 0 & 0 & 68,993\\
        $5\times10^{-4}$ & -0.732798 & 2 & 108,805 & 2 & 4 & 108,813\\
        $2\times10^{-4}$ & -0.752227 & 46 & 144,765 & 1,280 & 1,847 & 147,938\\
        $1\times10^{-4}$ & -0.757748 & 158 & 160,555 & 43,230 & 111,302 & 315,245\\ \hline
        0 (Full) & & 1,458 & 777,843 & $2\times10^{8}$ & $3\times10^{10}$ \\
        \hline
    \end{tabular}
    
    \caption{(Top) Correlation energy of stretched circumcoronene
    \\
    (Bottom) SCC amplitude count (in spin orbitals) and raw correlation energies
    }
    \label{fig:circum_stretch}
\end{figure}

We now study the circumcoronene system with C-C bonds stretched by 10\% from its optimized geometry. C-H bond distance was kept constant. We expect SCC to inherit single-reference CC methods’ weakness against stretched, strongly correlated systems and use this system to probe its properties. For this system, it is unclear whether CCSDT or high-bond-dimension DMRG results provide a more reliable estimate of the ground-state energy. Therefore, we present the raw correlation energies in Figure \ref{fig:circum_stretch}. For this system, we note that the SCC calculations for loose SCI thresholds did not converge despite trying out different convergence strategies. It is unclear whether this convergence issue would persist with a more robust implementation, and we intend to explore this in the future with a production-level SCC code. The three converged SCC data points lie close to and appear to converge to the high-bond-dimension DMRG and CCSDT results. From this exploratory study, it seems like SCC may remain a viable choice for systems with medium-strength static correlation once important amplitudes are included. One could follow up with, for example, a systematic study along bond dissociation coordinates to strengthen our understanding in this direction.

\subsection{Zinc Phthalocyanine}

\begin{figure}[tb]
    \centering
    \includegraphics[width=0.7\linewidth]{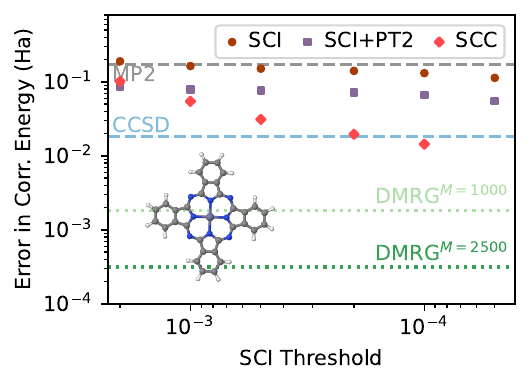}

    \resizebox{\linewidth}{!}{
    \setlength{\tabcolsep}{4pt}
    \begin{tabular}{c|c|*{6}{>{\centering\arraybackslash}p{1.6cm}}| >{\centering\arraybackslash}p{1.6cm}}
        \hline
        SCI & SCC Corr.& \multicolumn{7}{c}{Number of Amplitudes}\\ \cline{3-9}
        Threshold & Energy (Ha) & $T_1$ & $T_2$ & $T_3$ & $T_4$ & $T_5$ & $T_6$ & Total \\
        \hline
        $2\times10^{-3}$ & -0.368602 & 0 & 16,726 & 0 & 0 & 0 & 0 & 16,726\\
        $1\times10^{-3}$ & -0.415473 & 8 & 30,773 & 4 & 20 & 0 & 0 & 30,805\\
        $5\times10^{-4}$ & -0.438919 & 18 & 50,819 & 386 & 1,813 & 0 & 0 & 53,036\\
        $2\times10^{-4}$ & -0.450387 & 86 & 78,641 & 14,798 & 40,877 & 0 & 0 & 134,402\\
        $1\times10^{-4}$ & -0.455562 & 222 & 98,271 & 105,384 & 288,389 & 12 & 100 & 492,378\\ \hline
        0 (Full) & & 988 & 355,186 & $6\times10^{7}$ & $6\times10^{9}$ & $3\times10^{11}$ & $1\times10^{13}$\\
        \hline
    \end{tabular}
    }
    
    \caption{(Top) Error in correlation energy relative to CCSDT energy for ZnPc
    \\
    (Bottom) SCC amplitude count (in spin orbitals) and raw correlation energies}
    \label{fig:znpc}
\end{figure}

We use Zinc Phthalocyanine (ZnPc) as a representative to apply SCC on systems with interactions between the metal d-orbitals and the macrocyclic $\pi$-system. Similar to circumcoronene, we construct the active space using AVAS\cite{avas1,avas2} consisting of the carbon and nitrogen $2p_z$ orbitals and the zinc $3d$ orbitals. Hartree-Fock orbitals were evaluated at the def2-SVP basis set. We extrapolate the SCC correlation energy using a linear fit to $1 / N_\text{amplitudes}$ to obtain an estimate of the ground-state correlation energy, which is about 7 mHa away from the CCSDT correlation energy. Since CCSDT and DMRG ($M = 4000$) nearly match each other (0.3 mHa difference), we take the CCSDT energy as the best guess for the ground-state correlation energy. As shown in Figure \ref{fig:znpc}, benchmarks on the ZnPc system share similar observations: i) SCC exhibits robust, monotonic convergence with a tighter SCI threshold, ii) SCC outperforms SCI for a given SCI threshold, iii) SCC is more accurate than SCI+PT2 except for the loose SCI threshold regime, and iv) As shown in the Supporting Information, SCC calculation that optimizes amplitudes as many as CCSD shows comparable accuracy.

\subsection{Boron Cluster}

\begin{figure}[tb]
    \centering
    \includegraphics[width=0.7\linewidth]{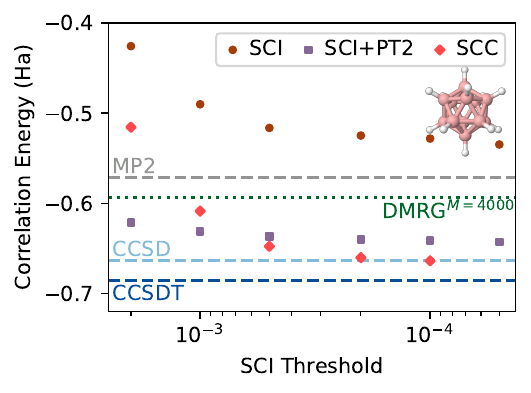}

    \setlength{\tabcolsep}{4pt}
    \begin{tabular}{c|c|*{4}{>{\centering\arraybackslash}p{1.6cm}}| >{\centering\arraybackslash}p{1.6cm}}
        \hline
        SCI & SCC Corr.& \multicolumn{5}{c}{Number of Amplitudes}\\ \cline{3-7}
        Threshold & Energy (Ha) & $T_1$ & $T_2$ & $T_3$ & $T_4$ & Total \\
        \hline
        $2\times10^{-3}$ & -0.515567 & 0 & 70,341 & 0 & 0 & 70,341\\
        $1\times10^{-3}$ & -0.608580 & 0 & 167,869 & 0 & 0 & 167,869\\
        $5\times10^{-4}$ & -0.647587 & 0 & 335,732 & 0 & 0 & 335,732\\
        $2\times10^{-4}$ & -0.660124 & 54 & 614,657 & 1,968 & 2,759 & 619,438\\
        $1\times10^{-4}$ & -0.663677 & 218 & 821,479 & 46,450 & 85,356 & 953,503\\ \hline
        0 (Full) & & 2,590 & $2\times10^{6}$ & $1\times10^{9}$ & $3\times10^{11}$ \\
        \hline
    \end{tabular}
    \caption{(Top) Correlation energy of boron cluster
    \\
    (Bottom) SCC amplitude count (in spin orbitals) and raw correlation energies}
    \label{fig:b12h122-}
\end{figure}

We now report benchmark results on boron cluster ($[\text{B}_{12}\text{H}_{12}]^{2-}$), a sample system with three-dimensional topology. Because the three-dimensional topology lacks a natural linear ordering, DMRG is expected to converge slowly with the bond dimension for this system. On the other hand, SCI is agnostic to the topology of the chemical system. Based on SCC’s improved accuracy over SCI and SCI+PT2 in the other examples, we model the boron cluster to assess whether SCC has the potential to improve studies of these high-dimensional chemical systems. As shown in Figure \ref{fig:b12h122-}, DMRG indeed seems to converge slowly with bond dimension. The large differences between CCSD, CCSD(T), and CCSDT correlation energies suggest the possibility of over-correlation in the CCSDT results, and the extrapolated SCC correlation energies indeed predict a correlation energy 20 mHa above CCSDT. Determining the \textit{best estimate} for the ground-state energy, however, is an abstruse matter due to the large spread of the calculated energies among the different methods, and we present the raw correlation energies in Figure \ref{fig:b12h122-}. Despite the limitations, SCC again outperforms SCI in all regimes and SCI+PT2 in the tighter SCI threshold. SCC, just like SCI and SCI+PT2, exhibits robust convergence with the SCI threshold, agnostic to the system topology, and further improves upon SCI and SCI+PT2.

\section{Discussions}\label{sec:discussion}

\paragraph{Computational Cost}
In the current implementation of SCC, RAM usage is dictated by the size of the $T$-amplitudes and the residual tensor. In the residual tensor evaluation step, each contributing term is contracted on-the-fly. Unlike modern dense CC implementations, which evaluate intermediates to reduce the computational cost scaling, our initial pilot code works with unfactorized terms without additional storage. At this stage, we have deliberately made this choice since further studies are necessary to assess whether the intermediate tensors have comparable levels of sparsity as the input $T$-amplitudes. Consequently, in terms of RAM usage, the SCI step is the bottleneck in the current SCC workflow. Even the large butane SCC calculation ($\varepsilon_\text{var} = 10^{-4}$, 707688 amplitudes) only uses a modest 4314 MB of RAM.

In terms of wall time required to run SCC calculations, the unfactorized terms leave room for future optimizations. For example, in dense CC implementations, evaluation of a residual that has a form
\begin{equation}
    R_{ij}^{ab} \leftarrow \frac{1}{16} \langle cd || kl \rangle T_{kl}^{ab} T_{ij}^{cd}
\end{equation}
avoids the $\mathcal{O}(N_o^4 N_v^4)$ cost scaling by introducing intermediate tensor.
\begin{align}
    W_{cd}^{ab} &= \langle cd || kl \rangle T_{kl}^{ab}\\
    R_{ij}^{ab} &= \frac{1}{16} W_{cd}^{ab} T_{ij}^{cd}
\end{align}
The two-step contraction now has a $\mathcal{O}(N_o^2 N_v^4)$ cost scaling, albeit with an intermediate storage requirement.\cite{cc_factor1,cc_factor2} It is likely that the optimal choice among different candidate intermediate tensor definitions or on-the-fly evaluation will depend on the sparsity structure of the $T$-amplitudes and the specific tensor contraction. We are actively exploring this direction to develop a production-level implementation of SCC. As is, SCC has a competitive wall-time-accuracy trade-off compared to high-bond-dimension DMRG for chemical structures with high-dimensional topology, as illustrated in the Supporting Information, but a future implementation may improve competitiveness for systems with simpler topology. We also comment that $T_1$-dressing of the Hamiltonians, coupled with density fitting for the electronic repulsion integrals, will likely accelerate SCC residual evaluations.\cite{cc_t1dress} Analogous to dense CC implementations, $T_1$-dressing would lead to a noticeable reduction in wall time without changing the asymptotic scaling.

\paragraph{Amplitude Selection}
In the present work, we offload amplitude selection to HCI, noting its success in identifying compact determinant subspaces for SCI calculations. Because the exponential ansatz in CC accounts for the contributions from the disconnected products of low-rank $T$-amplitudes, a good determinant selection scheme for SCI does not necessarily imply a good amplitude selection strategy. To assess the quality of SCI-based amplitude-space selection, we first quantify in Table \ref{table:sig_amps} the proportion of amplitudes included by the selection scheme but deemed insignificant after CC optimization.

\begin{table}[tb]
    \centering
    \begin{tabular}{|c|c|c|c|} \hline
        & $|t| < 10^{-8}$ & $|t| < 10^{-7}$  & $|t| < 10^{-6}$\\ \hline
        Circumcoronene ($\varepsilon_\text{var} = 10^{-4}$) & 0.074\% & 0.686\% & 6.51\% \\
        Stretched Circumcoronene ($\varepsilon_\text{var} = 10^{-4}$) & 0.034\% & 0.291\% & 3.05\% \\
        ZnPc ($\varepsilon_\text{var} = 10^{-4}$) & 0.261\% & 2.60\% & 19.2\%\\
        $[\text{B}_{12}\text{H}_{12}]^{2-}$ ($\varepsilon_\text{var} = 2\times10^{-4}$) & 0.003\% & 0.027\% & 0.192\% \\
        \hline
    \end{tabular}
    \caption{Proportion of non-significant amplitudes in the optimized SCC wavefunction}
    \label{table:sig_amps}
\end{table}

Results in Table \ref{table:sig_amps} suggest that most amplitudes chosen by SCI are significant after SCC optimization, with some variability across systems. In the cluster decomposition step of the SCI wavefunction, we threshold the $T$-amplitudes at $10^{-8}$, and the same threshold identifies the significant CC amplitudes. The results indicate that the SCI-based selection scheme does not erroneously include many insignificant amplitudes.

In Table \ref{table:loose_tight}, we analyze how well the SCI-based scheme captures the leading $N$ amplitudes. In other words, we quantify the true positives. Precise analysis of the true positives requires knowledge of the Full CC wavefunction, which we do not have on hand. We therefore use the SCC wavefunction with the tightest threshold as a proxy and check how many of the most significant $N_\text{loose}$ amplitudes in the tightest SCC wavefunction the SCI-based selection scheme chose for the loose threshold calculation. Across diverse systems, we observe that the SCI-based selection scheme identifies roughly 80-90\% of the most significant amplitudes.

\begin{table}[tb]
    \centering
    \begin{tabular}{|c|c|c|c|c|} \hline
        \multirow{2}{*}{Reference SCC Wavefunction} & \multicolumn{4}{c|}{$\varepsilon_\text{var}$}\\ \cline{2-5}
        & $2\times10^{-4}$ & $5\times10^{-4}$  & $1\times10^{-3}$ & $2\times10^{-3}$\\ \hline
        Circumcoronene ($\varepsilon_\text{var} = 10^{-4}$) & 92.0\% & 91.7\% & 87.7\% & 78.1\% \\
        Stretched Circumcoronene ($\varepsilon_\text{var} = 10^{-4}$) & 91.1\% & 90.4\% & 85.6\% & 75.8\% \\
        ZnPc ($\varepsilon_\text{var} = 10^{-4}$) & 65.7\% & 86.1\% & 86.2\% & 81.7\%\\
        $[\text{B}_{12}\text{H}_{12}]^{2-}$ ($\varepsilon_\text{var} = 2\times10^{-4}$) & N/A & 78.8\% & 83.0\% & 83.4\% \\
        \hline
    \end{tabular}
    \caption{Proportion of significant amplitudes captured in looser SCI thresholds}
    \label{table:loose_tight}
\end{table}

This result implies that the SCI-based selection scheme is remarkably close to optimal. The true positive rate of 80-90\% indicates a false negative rate of 10-20\%. From the $M$ astronomically large number of possible full CC amplitudes, the false positive rate measures the proportion of the $M-N$ insignificant amplitudes the SCI-based selection scheme included. Because $M\gg N$ and the number of false positives is relatively small in Table \ref{table:loose_tight}, the false positive rate is insignificantly small. This indicates that there is at most 10-20\% room for improvement for future amplitude selection schemes. While a computationally cheaper selection scheme can be useful, the SCI-based selection scheme is very robust in terms of target accuracy.

\paragraph{Perturbative Corrections}
SCI methods significantly benefit from perturbative corrections, both for accurate energies and for robust extrapolative schemes. In particular, extrapolating SCI+PT2 energies to the limit where perturbative corrections vanish has gained popularity due to its effectiveness and a logical rationale for the form of the fitting function.\cite{sci_fitting} We plan to explore an analogous non-iterative correction for SCC in the near future. We envision that the perturbative correction will resemble the treatment for triples in CCSD(T) or quadruples in CCSDT(Q), approximating the contributions from external amplitudes linked to internal amplitudes.\cite{ccsd_t,ccsdt_q} Another avenue worth exploring is a CC2- or CC3-like approximate treatment of higher-rank operators.\cite{cc_2,cc_3}

\section{Conclusions}\label{sec:conclusion}

We introduce a novel formulation of arbitrary-order Selected Coupled Cluster (SCC) that uses SCI wavefunctions to guide the choice of amplitude subspaces. Through benchmark calculations on several chemical systems, we have shown that: 1) SCC correlation energies are accurate and converge rapidly, outperforming SCI calculations with an identical threshold $\varepsilon_\text{var}$ and SCI+PT2 in the tight-threshold regime; 2) SCC maintains robust performance even with chemical systems of high-dimensional topology, where DMRG typically requires a high bond dimension for convergence; and 3) SCC is better or competitive to CCSD with a comparable number of amplitudes. We analyzed the computational cost of SCC calculations, assessed the amplitude selection criteria based on Heat-bath CI, and commented on a potential perturbative correction scheme. We conclude that SCC is an efficient drop-in improvement strategy to follow after running modern SCI calculations.

The current pilot implementation of SCC limits the number of $T$-amplitudes we can treat on a single compute node to about 1M amplitudes. This suggests an improved implementation as a natural subsequent direction we plan to investigate. A production-level implementation will consider the factorization techniques that benefit dense CC code to accelerate the residual evaluation step. We also plan to explore multi-node MPI parallelism to expand the subspace size to reach what has been reported in recent Heat-bath CI implementations. \cite{hci_modern} Our analysis of false positives and negatives indicates that the modern SCI-based selection scheme robustly finds the significant CC amplitudes. It leaves little margin for improvement by any alternative scheme, though a computationally cheaper selection scheme can be useful. Finally, adding a non-iterative perturbative correction will offer a pathway to energy extrapolation, establishing SCC as a robust workflow for accurate and efficient evaluation of electronic energies.

\begin{acknowledgement}
This work was supported by the U.S.
Department of Energy, Office of Science, National Quantum Information Science Research Centers, Co-design Center for Quantum Advantage (C2QA) under Contract No. DE-SC0012704 (PNNL FWP 76274).

Authors thank Hironori Kondo, Lea Northcote, Shaun Weatherly, Noah Whelpley, Abigail McClain Gomez, William Kirby, Mario Motta, and Javier Robledo-Moreno for the insightful discussions. Jmol was used to generate the ball-and-stick diagrams.\cite{jmol}
\end{acknowledgement}

\begin{suppinfo}
\begin{itemize}
    \item Additional data and Molecular structures in xyz format (pdf)
\end{itemize}
\end{suppinfo}


\clearpage 
\bibliography{bib}

@article{wicked,
  author  = {Evangelista, Francesco A.},
  title   = {Automatic derivation of many-body theories based on general {Fermi} vacua},
  journal = {J. Chem. Phys.},
  volume  = {157},
  pages   = {064111},
  year    = {2022},
  doi     = {10.1063/5.0097858},
  url     = {https://doi.org/10.1063/5.0097858},
}

@article{clusterdec,
  author  = {Susi Lehtola and Norm M. Tubman and K. Birgitta Whaley and Martin Head-Gordon
},
  title   = {Cluster decomposition of full configuration interaction wave functions: A tool for chemical interpretation of systems with strong correlation},
  journal = {J. Chem. Phys.},
  volume  = {147},
  pages   = {154105},
  year    = {2017},
  doi     = {10.1063/1.4996044},
  url     = {https://doi.org/10.1063/1.4996044},
}

@article{pyscf,
	title        = {Recent developments in the PySCF program package},
	author       = {Sun, Qiming and Zhang, Xing and Banerjee, Samragni and Bao, Peng and Barbry, Marc and Blunt, Nick S. and Bogdanov, Nikolay A. and Booth and Cui, Zhi-Hao and Eriksen, Janus J. and Gao, Yang and Guo, Sheng and Hermann, Jan and Hermes, Matthew R. and Koh, Kevin and Koval and Li, Zhendong and Liu, Junzi and Mardirossian and Motta, Mario and Mussard, Bastien and Pham and Purwanto, Wirawan and Robinson, Paul J. and Ronca, Enrico and Sayfutyarova, Elvira R. and Scheurer, Maximilian and Schurkus, Henry F. and Smith, James E. T. and Sun, Chong and Sun, Shi-Ning and Upadhyay, Shiv and Wagner, Lucas K. and Wang, Xiao and White, Alec and Whitfield, James Daniel and Williamson, Mark J. and Wouters, Sebastian and Yang, Jun and Yu, Jason M. and Zhu, Tianyu and Berkelbach, Timothy C. and Sharma and Chan, Garnet Kin-Lic},
	year         = 2020,
	month        = {07},
	journal      = {The Journal of Chemical Physics},
	volume       = 153,
	number       = 2,
	pages        = {024109},
	doi          = {10.1063/5.0006074},
	issn         = {0021-9606},
	url          = {https://doi.org/10.1063/5.0006074},
	eprint       = {https://pubs.aip.org/aip/jcp/article-pdf/doi/10.1063/5.0006074/16722275/024109\_1\_online.pdf}
}

@misc{jmol,
	title        = {Jmol: an open-source Java viewer for chemical structures in 3D.},
	author       = {{Jmol Developers}},
	url          = {http://www.jmol.org}
}

@article{arrow1,
Author = {Junhao Li and  Matthew Otten and Adam A. Holmes and Sandeep Sharma and C. J. Umrigar},
Title = {Fast Semistochastic Heat-Bath Configuration Interaction},
Journal = {J. Chem. Phys.},
Year = {2018},
Volume = {148},
Pages = {214110}
}

@article{arrow2,
Author = {Sandeep Sharma and Adam A. Holmes and Guillaume Jeanmairet and Ali Alavi and C. J. Umrigar},
Title = {Semistochastic Heat-Bath Configuration Interaction Method: Selected
   Configuration Interaction with Semistochastic Perturbation Theory},
Journal = {J. Chem. Theory Comput.},
Year = {2017},
Volume = {13},
Pages = {1595-1604},
DOI = {10.1021/acs.jctc.6b01028},
}

@article{arrow3,
Author = {Adam A. Holmes and Norm M. Tubman and C. J. Umrigar},
Title = {Heat-bath Configuration Interaction: An efficient selected CI algorithm inspired by heat-bath sampling},
Journal = {J. Chem. Theory Comput.},
Volume = {12},
Pages = {3674-3680},
Year = {2016}
}

@article{block2-1,
  author  = {Zhai, Huanchen and Larsson, Henrik R. and Lee, Sunghoon and Cui, Zhi-Hao and Zhu, Tianyu and Sun, Changsu and Peng, Linus and Peng, Ruojie and Liao, King and T{\"o}lle, J{\"o}rg and Yang, J{\"u}rg and Li, Shichuan and Chan, Garnet Kin-Lic},
  title   = {Block2: A comprehensive open source framework to develop and apply state-of-the-art DMRG algorithms in electronic structure and beyond},
  journal = {The Journal of Chemical Physics},
  volume  = {159},
  number  = {23},
  pages   = {234801},
  year    = {2023},
  doi     = {10.1063/5.0180424}
}

@article{block2-2,
  author  = {Chan, Garnet Kin-Lic and Head-Gordon, Martin},
  title   = {Highly correlated calculations with a polynomial cost algorithm: A study of the density matrix renormalization group},
  journal = {The Journal of Chemical Physics},
  volume  = {116},
  number  = {11},
  pages   = {4462--4476},
  year    = {2002},
  doi     = {10.1063/1.1449459}
}

@article{block2-3,
  author  = {Sharma, Sandeep and Chan, Garnet Kin-Lic},
  title   = {Spin-adapted density matrix renormalization group algorithms for quantum chemistry},
  journal = {The Journal of Chemical Physics},
  volume  = {136},
  number  = {12},
  pages   = {124121},
  year    = {2012},
  doi     = {10.1063/1.3695642}
}

@misc{qscc_repo,
  author       = {Cho, Minsik},
  title        = {Selected Coupled Cluster},
  year         = {2026},
  howpublished = {\url{https://github.com/troyvvgroup/qscc}},
  note         = {Accessed: July 6, 2026}
}

@article{avas1,
  author  = {Elvira R. Sayfutyarova and Qiming Sun and Garnet Kin-Lic Chan and Gerald Knizia},
  title   = {Automated Construction of Molecular Active Spaces from Atomic Valence Orbitals},
  journal = {J. Chem. Theory Comput.},
  volume  = {13},
  number  = {9},
  pages   = {4063--4078},
  year    = {2017},
  doi     = {10.1021/acs.jctc.7b00128}
}

@article{avas2,
  author  = {Elvira R. Sayfutyarova and Sharon Hammes-Schiffer},
  title   = {Constructing Molecular $\pi$-Orbital Active Spaces for Multireference Calculations of Conjugated Systems},
  journal = {J. Chem. Theory Comput.},
  volume  = {15},
  number  = {3},
  pages   = {1679--1689},
  year    = {2019},
  doi     = {10.1021/acs.jctc.8b01196}
}

@article{cipsi,
  author  = {B. Huron and J. P. Malrieu and P. Rancurel},
  title   = {Iterative perturbation calculations of ground and excited state energies from multiconfigurational zeroth‐order wavefunctions},
  journal = {J. Chem. Phys.},
  volume  = {58},
  number  = {12},
  pages   = {5745--5759},
  year    = {1973},
  doi     = {10.1063/1.1679199}
}

@article{fciqmc,
  author  = {George Booth and Alex J. W. Thom and Ali Alavi},
  title   = {Fermion Monte Carlo without fixed nodes: A game of life, death, and annihilation in Slater determinant space},
  journal = {J. Chem. Phys.},
  volume  = {131},
  pages   = {054106},
  year    = {2009},
  doi     = {10.1063/1.3193710}
}

@article{aci,
  author  = {Jeffrey B. Schriber and Francesco A. Evangelista},
  title   = {Adaptive Configuration Interaction for Computing Challenging Electronic Excited States with Tunable Accuracy},
  journal = {J. Chem. Theory Comput.},
  volume  = {13},
  number  = {11},
  pages   = {5354-5366},
  year    = {2017},
  doi     = {10.1021/acs.jctc.7b00725}
}

@article{asci,
  author  = {Norm M. Tubman and Joonho Lee and Tyler Y. Takeshita and Martin Head-Gordon and K. Birgitta Whaley},
  title   = {A deterministic alternative to the full configuration interaction quantum Monte Carlo method},
  journal = {J. Chem. Phys.},
  volume  = {145},
  pages   = {044112},
  year    = {2016},
  doi     = {10.1063/1.4955109}
}

@article{tcc-casci,
  author  = {Tomoko Kinoshita and Osamu Hino and Rodney J. Bartlett},
  title   = {Coupled-cluster method tailored by configuration interaction},
  journal = {J. Chem. Phys.},
  volume  = {123},
  pages   = {074106},
  year    = {2005},
  doi     = {10.1063/1.2000251}
}

@article{tcc-dmrg,
  author  = {Veis, Libor and Antal{\'{i}}k, Andrej and Brabec, Ji{\v{r}}{\'{i}} and Neese, Frank and Legeza, {\"{O}}rs and Pittner, Ji{\v{r}}{\'{i}}},
  title   = {Coupled Cluster Method with Single and Double Excitations Tailored by Matrix Product State Wave Functions},
  journal = {J. Phys. Chem. Lett.},
  volume  = {7},
  number  = {20},
  pages   = {4072--4078},
  year    = {2016},
  doi     = {10.1021/acs.jpclett.6b01908}
}

@article{tcc-benchmark,
  author  = {Maximilian Mörchen and Leon Freitag and Markus Reiher},
  title   = {Tailored coupled cluster theory in varying correlation regimes},
  journal = {J. Chem. Phys.},
  volume  = {153},
  pages   = {244113},
  year    = {2020},
  doi     = {10.1063/5.0032661}
}

@article{ccpq1,
  author  = {Nicholas Bauman and Jun Shen and Piotr Piecuch},
  title   = {Combining active-space coupled-cluster approaches with moment energy corrections via the CC(P;Q) methodology: connected quadruple excitations},
  journal = {Mol. Phys.},
  volume  = {115},
  pages   = {2860--2891},
  year    = {2017},
  doi     = {10.1080/00268976.2017.1350291}
}

@article{ccpq2,
  author  = {Jun Shen and Piotr Piecuch},
  title   = {Combining active-space coupled-cluster methods with moment energy corrections via the CC(P;Q) methodology, with benchmark calculations for biradical transition states},
  journal = {J. Chem. Phys.},
  volume  = {136},
  pages   = {144104},
  year    = {2012},
  doi     = {10.1063/1.3700802}
}

@article{ccpq3,
  author  = {Jun Shen and Piotr Piecuch},
  title   = {Merging Active-Space and Renormalized Coupled-Cluster Methods via the CC(P;Q) Formalism, with Benchmark Calculations for Singlet–Triplet Gaps in Biradical Systems},
  journal = {J. Chem. Theory Comput.},
  volume  = {8},
  number  = {12},
  pages   = {4968–4988},
  year    = {2012},
  doi     = {10.1021/ct300762m}
}

@article{ccpq4,
  author  = {J. Emiliano Deustua and Jun Shen and Piotr Piecuch},
  title   = {Converging High-Level Coupled-Cluster Energetics by Monte Carlo Sampling and Moment Expansions},
  journal = {Phys. Rev. Lett.},
  volume  = {119},
  pages   = {223003},
  year    = {2017},
  doi     = {10.1103/PhysRevLett.119.223003}
}

@article{atcc,
  author  = {Dmitry I. Lyakh and Rodney J. Bartlett},
  title   = {An adaptive coupled-cluster theory: @CC approach},
  journal = {J. Chem. Phys.},
  volume  = {133},
  pages   = {244112},
  year    = {2010},
  doi     = {10.1063/1.3515476}
}

@article{fccr,
  author  = {Enhua Xu and Motoyuki Uejima and Seiichiro Lenka Ten-no},
  title   = {Full Coupled-Cluster Reduction for Accurate Description of Strong Electron Correlation},
  journal = {Phys. Rev. Lett.},
  volume  = {121},
  pages   = {113001},
  year    = {2018},
  doi     = {10.1103/PhysRevLett.121.113001}
}

@article{ccmc,
  author  = {Alex J. W. Thom},
  title   = {Stochastic Coupled Cluster Theory},
  journal = {Phys. Rev. Lett.},
  volume  = {105},
  pages   = {263004},
  year    = {2010},
  doi     = {10.1103/PhysRevLett.105.263004}
}

@article{uscc,
  author  = {Dmitry A. Fedorov and Yuri Alexeev and Stephen K. Gray and Matthew Otten},
  title   = {Unitary Selective Coupled-Cluster Method},
  journal = {Quantum},
  volume  = {6},
  pages   = {703},
  year    = {2022},
  doi     = {10.22331/q-2022-05-02-703}
}

@article{wicktheorem,
  author  = {Gian Carlo Wick},
  title   = {The Evaluation of the Collision Matrix},
  journal = {Phys. Rev.},
  volume  = {80},
  pages   = {268},
  year    = {1950},
  doi     = {10.1103/PhysRev.80.268}
}

@article{cc1,
  author  = {Ji{\v{r}}{\'{i}} {\v{C}}{\'{i}}{\v{z}}ek},
  title   = {On the Correlation Problem in Atomic and Molecular Systems. Calculation of Wavefunction Components in Ursell‐Type Expansion Using Quantum‐Field Theoretical Methods},
  journal = {J. Chem. Phys.},
  volume  = {45},
  pages   = {4256--4266},
  year    = {1966},
  doi     = {10.1063/1.1727484}
}

@inbook{cc2,
    author = {Čížek, Jiří},
    publisher = {John Wiley \& Sons, Ltd},
    isbn = {9780470143599},
    title = {On the Use of the Cluster Expansion and the Technique of Diagrams in Calculations of Correlation Effects in Atoms and Molecules},
    booktitle = {Advances in Chemical Physics},
    chapter = {},
    pages = {35-89},
    doi = {https://doi.org/10.1002/9780470143599.ch2},
    year = {1969},
}

@article{cc3,
  author  = {Ji{\v{r}}{\'{i}} {\v{C}}{\'{i}}{\v{z}}ek and J. Paldus},
  title   = {Correlation problems in atomic and molecular systems III. Rederivation of the coupled-pair many-electron theory using the traditional quantum chemical methods},
  journal = {International Journal of Quantum Chemistry},
  volume  = {5},
  number  = {4},
  pages   = {359--379},
  year    = {1971},
  doi     = {10.1002/qua.560050402}
}

@article{cc4,
  author  = {Rodney J. Bartlett and Monika Musial},
  title   = {Coupled-cluster theory in quantum chemistry},
  journal = {Rev. Mod. Phys.},
  volume  = {79},
  pages   = {291},
  year    = {2007},
  doi     = {10.1103/RevModPhys.79.291}
}

@article{ccsd_t1,
title = {Non-iterative fifth-order triple and quadruple excitation energy corrections in correlated methods},
journal = {Chemical Physics Letters},
volume = {165},
number = {6},
pages = {513-522},
year = {1990},
doi = {10.1016/0009-2614(90)87031-L},
author = {Rodney J. Bartlett and J.D. Watts and S.A. Kucharski and J. Noga},
}

@article{ccsd_t2,
title = {A fifth-order perturbation comparison of electron correlation theories},
journal = {Chemical Physics Letters},
volume = {157},
number = {6},
pages = {479-483},
year = {1989},
doi = {10.1016/S0009-2614(89)87395-6},
author = {Krishnan Raghavachari and Gary W. Trucks and John A. Pople and Martin Head-Gordon},
}

@article{cc_factor1,
  author  = {Gustavo E. Scuseria and Curtis L. Janssen and Henry F. Schaefer, III},
  title   = {An efficient reformulation of the closed‐shell coupled cluster single and double excitation (CCSD) equations},
  journal = {J. Chem. Phys.},
  volume  = {89},
  pages   = {7382--7387},
  year    = {1988},
  doi     = {10.1063/1.455269}
}

@article{cc_factor2,
  author  = {John F. Stanton and J\"{u}rgen Gauss and John D. Watts and Rodney J. Bartlett},
  title   = {A direct product decomposition approach for symmetry exploitation in many‐body methods. I. Energy calculations},
  journal = {J. Chem. Phys.},
  volume  = {94},
  pages   = {4334--4345},
  year    = {1991},
  doi     = {10.1063/1.460620}
}

@article{sci_fitting,
  author  = {Burton, Hugh G. A. and Loos, Pierre-Fran{\c{c}}ois},
  title   = {Rationale for the extrapolation procedure in selected configuration interaction},
  journal = {J. Chem. Phys.},
  volume  = {160},
  pages   = {104102},
  year    = {2024},
  doi     = {10.1063/5.0192458}
}

@article{ccsd_t,
title = {A fifth-order perturbation comparison of electron correlation theories},
journal = {Chemical Physics Letters},
volume = {157},
number = {6},
pages = {479-483},
year = {1989},
doi = {10.1016/S0009-2614(89)87395-6},
author = {Krishnan Raghavachari and Gary W. Trucks and John A. Pople and Martin Head-Gordon},
}

@article{ccsdt_q,
  author = {Bomble, Yannick J. and Stanton, John F. and K{\'a}llay, Mih{\'a}ly and Gauss, J{\"u}rgen},
  title   = {Coupled-cluster methods including noniterative corrections for quadruple excitations},
  journal = {J. Chem. Phys.},
  volume  = {123},
  pages   = {054101},
  year    = {2005},
  doi     = {10.1063/1.1950567}
}

@article{cc_2,
  author = {Ove Christiansen and Henrik Koch and Poul J{\o}rgensen},
  title   = {The second-order approximate coupled cluster singles and doubles model CC2},
  journal = {Chem. Phys. Lett.},
  volume  = {243},
  number  = {5-6},
  pages   = {409-418},
  year    = {1995},
  doi     = {10.1016/0009-2614(95)00841-Q}
}

@article{cc_3,
  author = {Henrik Koch and Ove Christiansen and Poul J{\o}rgensen and Alfredo M. Sanchez de Mer\'{a}s and Trygve Helgaker},
  title   = {The CC3 model: An iterative coupled cluster approach including connected triples},
  journal = {J. Chem. Phys.},
  volume  = {106},
  pages   = {1808-1818},
  year    = {1997},
  doi     = {10.1063/1.473322}
}

@article{hci_modern,
  author = {Duy-Khoi Dang and Joshua A. Kammeraad and Paul M. Zimmerman},
  title   = {Advances in Parallel Heat Bath Configuration Interaction},
  journal = {J. Phys. Chem. A},
  volume  = {127},
  number  = {1},
  pages   = {400--411},
  year    = {2023},
  doi     = {10.1021/acs.jpca.2c07949}
}

@article{cc_t1dress,
  author = {A. Eugene DePrince, III and C. David Sherrill},
  title   = {Accuracy and Efficiency of Coupled-Cluster Theory Using Density Fitting/Cholesky Decomposition, Frozen Natural Orbitals, and a t1‑Transformed Hamiltonian},
  journal = {J. Chem. Theory Comput.},
  volume  = {9},
  number  = {6},
  pages   = {2687--2696},
  year    = {2013},
  doi     = {10.1021/ct400250u}
}

@article{ccsdt,
  author = {Jozef Noga and Rodney J. Bartlett},
  title   = {The full CCSDT model for molecular electronic structure},
  journal = {J. Chem. Phys.},
  volume  = {86},
  pages   = {7041--7050},
  year    = {1987},
  doi     = {10.1063/1.452353}
}

@article{ccsdtq,
  author = {Stanislaw A. Kucharski and Rodney J. Bartlett},
  title   = {The coupled‐cluster single, double, triple, and quadruple excitation method},
  journal = {J. Chem. Phys.},
  volume  = {97},
  pages   = {4282--4288},
  year    = {1992},
  doi     = {10.1063/1.463930}
}

@article{dlpno_ccsd,
  author = {Christoph Riplinger and Frank Neese},
  title   = {An efficient and near linear scaling pair natural orbital based local coupled cluster method},
  journal = {J. Chem. Phys.},
  volume  = {138},
  pages   = {034106},
  year    = {2013},
  doi     = {10.1063/1.4773581}
}

@article{lpno_ccsd,
  author = {Frank Neese and Andreas Hansen and Dimitrios G. Liakos},
  title   = {Efficient and accurate approximations to the local coupled cluster singles doubles method using a truncated pair natural orbital basis},
  journal = {J. Chem. Phys.},
  volume  = {131},
  pages   = {064103},
  year    = {2009},
  doi     = {10.1063/1.3173827}
}

@article{cc_review,
  author = {Dmitry I. Lyakh and Monika Musial and Victor F. Lotrich and Rodney J. Bartlett},
  title   = {Multireference Nature of Chemistry: The Coupled-Cluster View},
  journal = {Chem. Rev.},
  volume  = {112},
  number  = {1},
  pages   = {182--243},
  year    = {2012},
  doi     = {10.1021/cr2001417}
}

@misc{pyscf_ccsdtq,
  title  = {High-performance parallel implementation of high-order coupled-cluster theories}, 
  author = {Yu Jin and Christopher Hillenbrand and Timothy C. Berkelbach and Huanchen Zhai},
  year   = {2026},
  eprint = {2607.00981},
  archivePrefix = {arXiv},
  primaryClass = {physics.chem-ph},
  url = {https://arxiv.org/abs/2607.00981}
}

\end{document}